\documentclass[a4paper]{jpconf}

\usepackage{graphicx}

\usepackage{amsmath,cleveref,amssymb,amsfonts,multirow,color,subfigure,textcomp,mathtools}

\usepackage{srcltx,soul,subfigure,bbm,bm}

\usepackage{threeparttable}

\newcommand{\bee}{\begin{equation}}
\newcommand{\ee}{\end{equation}}
\newcommand{\bma}{\begin{pmatrix}}
\newcommand{\ema}{\end{pmatrix}}
\newcommand{\balig}{\begin{align}}
\newcommand{\ealig}{\end{align}}

\newcommand{\bZ}{\mathbb{Z}}

\newcommand{\ba}{\begin{eqnarray}}
\newcommand{\ea}{\end{eqnarray}}
\newcommand{\ignore}[1]{}

\newcommand{\tix}{\tau_x}
\newcommand{\tiy}{\tau_y}
\newcommand{\tiz}{\tau_z}

\newcommand{\bk}{{\bf k}}
\newcommand{\ck}{\mathcal{K}}

\begin{document}
\title{Classification of crystalline topological semimetals with an application to Na$_3$Bi}

\author{Ching-Kai Chiu}

\address{Department of Physics and Astronomy, University of British Columbia, Vancouver, BC, Canada V6T 1Z1}
\address{Quantum Matter Institute, University of British Columbia, Vancouver BC, Canada V6T 1Z4}
\ead{chiu7@phas.ubc.ca}

\author{Andreas P. Schnyder}
\address{Max-Planck-Institut f\"ur Festk\"orperforschung,
  Hei\ss{}enbergstrasse 1, D-70569 Stuttgart, Germany} 
\ead{a.schnyder@fkf.mpg.de}

\begin{abstract}
Topological phases can not only be protected by internal symmetries (e.g., time-reversal symmetry),
but also by crystalline symmetries, such as reflection or rotation symmetry. Recently a complete topological classification of reflection symmetric insulators, superconductors,
nodal semimetals, and nodal superconductors has been established. 
In this article, after a brief review of the classification of reflection-symmetry-protected semimetals and nodal superconductors, we
discuss an example of a three-dimensional topological Dirac semimetal, which exhibits time-reversal symmetry as well as reflection and rotation symmetries. 
We compute the surface state spectrum of this Dirac semimetal and identify the crystal lattice symmetries that lead to the protection of the surface states.
We discuss the implications of our findings for the stability of the Fermi arc surface states of the Dirac material Na$_3$Bi.
Our analysis suggests that the Fermi arc of Na$_3$Bi is gapped except at time-reversal invariant surface momenta, which is in agreement with recent photoemission measurements.
\end{abstract}

\section{Introduction}

Finding new topological materials beyond the Bi-based topological insulators has become an important goal in solid state research.
In the past few years, it has been realized that not only internal symmetries~\cite{Schnyder2008gf,Kitaev,SchnyderAIP,Ryu2010ten,Chiu_nontrivial_surface}, such as time-reversal, but also crystal symmetries, for example crystal reflection or rotation, can
lead to the protection of topological states~\cite{teoPRB08,Fu2011uq,slaberNatPhys13,uenoPRL13,zhangPRL13,benalcazar2013,Teo_hughes_2013,Turner:2012bh,HughesPRB11,chiuPRB13,morimotoPRB13,Sato_Crystalline_PRB14}. In fact, a complete classification of reflection-symmetry-protected insulators and fully gapped superconductors
was recently presented in Refs.~\cite{chiuPRB13,morimotoPRB13,Sato_Crystalline_PRB14}.
The topological character of these crystalline topological insulators and superconductors gives rise to gapless states at the boundary.
Importantly, since the surface of a material has lower crystal symmetry than its bulk,
not all surfaces of a crystalline topological insulator (or superconductor) exhibit gapless boundary states.
Only those surfaces that are invariant under the crystal symmetry can support topological surface states.
Experimentally, SnTe has been identified to be a crystalline topological insulator with Dirac cone surface states, that are protected against gap opening by
reflection symmetry together with time-reversal symmetry~\cite{Tanaka:2012fk,Hsieh:2012fk,Xu2012,Dziawa2012uq_short}. Other candidate materials for reflection-symmetry-protected topological insulators are the anti-perovskites  Ca$_3$PbO and Sr$_3$PbO~\cite{OgataJPSJ2011,Kariyado_Ca3PbO,hsieFuAntiPero14}.

The concept of topological band theory can also be applied to nodal systems, such as semimetals and nodal superconductors. The stability of the Fermi surfaces (superconducting nodes)
of these materials is ensured by the conservation of a topological invariant which is defined in terms of an integral along a contour enclosing (encircling) the 
Fermi point (Fermi line). Topological nodal systems can be protected by global symmetries~\cite{WanVishwanathSavrasovPRB11,SchnyderRyuFlat,BrydonSchnyderTimmFlat,Brydon10,matsuuraNJP13,ZhaoWangPRL13,ZhaoWangPRB14,turnerVishwanathReview} (i.e., internal symmetries, such as time-reversal) 
as well as crystal lattice symmetries, or a combination of the two~\cite{ChiuSchnyder14,YangNagaosaNatComm14,morimotoFurusakiPRB14,poYao14}.
The nontrivial topology of the electronic band structure of these gapless materials manifests itself at the boundary in terms of protected surface states.
Depending on the symmetry properties and the dimension of the Fermi surface, these surface states 
form Dirac cones, flat bands, or Fermi arc states~\cite{matsuuraNJP13}.
Recently, several materials have been experimentally identified to be crystal-symmetry-protected topological semimetals.
 Among them are the Dirac materials Cd$_3$As$_2$~\cite{Yazdani_CdAs,Dirac_semimetal_Xi_Dai,neupaneDiracHasan,borisenkoPRLCd3As2,Cd3As2Chen2014,heLiCd3AsTransport,liangOngTransportCd3As2}
 and Na$_3$Bi~\cite{Liu21022014,Dai_predition_Na3Bi,xuLiuHasanArxiv13,Xu18122014}, whose Dirac points are protected by rotation symmetry. 
 The Fermi arc states of Na$_3$Bi have recently been observed in angle-resolved photoemission measurements~\cite{xuLiuHasanArxiv13,Xu18122014}.

In this paper, we first present in Sec.~\ref{reviewClassify} a short review of the topological classification of reflection symmetric semimetals and nodal superconductors~\cite{ChiuSchnyder14}. 
For brevity, we focus on the case where the Fermi surfaces (superconducting nodes) are located within the reflection plane. 
Second, we discuss in Sec.~\ref{secDiracSemimetal} a four-band tight-binding model on the cubic lattice describing a three-dimensional Dirac semimetal, which 
is protected by $C_4$ rotation symmetry. We discuss the surface states of this topological semimetal and determine the symmetries that protect the gapless surface states.
Our tight-binding model has similar symmetry properties and topological features as the Dirac semimetal Na$_3$Bi.
Hence, from our topological analysis we can obtain information about the stability of the Fermi arc surface states of Na$_3$Bi, see Sec.~\ref{secImplicationNa3Bi}.

\section{Topological classification of reflection symmetric semimetals and nodal superconductors}
\label{reviewClassify}

Before discussing a concrete example of a crystalline topological semimetal, let us briefly review the topological classification of reflection symmetric semimetals
and nodal superconductors~\cite{ChiuSchnyder14}.
The topological properties of reflection symmetric semimetals (nodal superconductors) can be classified in a similar manner
as those of reflection symmetric insulators  (fully gapped superconductors)~\cite{chiuPRB13,morimotoPRB13,Sato_Crystalline_PRB14}.
The classification depends on the internal symmetry properties of the nodal system as well as 
on whether the reflection symmetry commutes or anticommutes with the internal symmetries. 
There are three fundamental internal symmetries, which 
act locally in position space, namely time-reversal symmetry (TRS), particle-hole symmetry (PHS), and chiral or sublattice symmetry (SLS).
In momentum space, time-reversal and particle-hole symmetry act on the Bloch Hamiltonian $H({\bf k})$ as
\begin{subequations}
\begin{equation} \label{eqTRS}
T^{-1} H ( - {\bf k} ) T = + H ( {\bf k} )
\end{equation}
and
\begin{equation}
C^{-1} H ( - {\bf k} ) C = - H ( {\bf k} ) ,
\end{equation}
\end{subequations}
respectively, where $T = \mathcal{K}U_T$ and $C = \mathcal{K}U_C$ are antiunitary operators and 
$\mathcal{K}$ is the complex conjugation operator.
Chiral symmetry is implemented in terms of a unitary matrix $S$ which anti-commutes with the Hamiltonian, 
i.e., $S H ( {\bf k} ) + H ( {\bf k}Ê) S$ = 0.

Crystal reflection symmetry acts non-locally in position space. In the following, we consider a $d_{\mathrm{BZ}}$-dimensional 
Hamiltonian $H({\bf k})$ which is symmetric under reflection in the first direction. 
Hence, $H ({\bf k})$ satisfies
\begin{eqnarray}
R^{-1} H ( -k_1, \tilde{\bf k} ) R
=
H ( k_1, \tilde{\bf k} ) ,
\end{eqnarray}
where $\tilde{\bf k} = (k_2, k_3, \ldots, k_{d_{\mathrm{BZ}} })$ and $R$ is a unitary reflection operator. 
Since a phase factor can be absorbed in the definition of the electron creation and annihilation operators, we can assume that $R$ is  
Hermitian, i.e., $R^{\dag} = R$. 
As a consequence, 
the eigenvalues of $R$ are either $+1$ or $-1$ and all
commutation and anti-commutation relations between $R$ and the internal symmetries are defined unambiguously, i.e.,
\begin{eqnarray}
T R T^{-1} = \eta_T R, \quad
C R C^{-1} = \eta_C R \quad
\textrm{and} \quad
S R S^{-1} = \eta_S R,  
\end{eqnarray}
where the three indices $\eta_T$,  $\eta_C$, and   $\eta_S$ specify whether $R$ commutes ($+1$) or anti-commutes ($-1$) with the corresponding internal symmetry operator. For the symmetry classes AI, AII, AIII, C, and D these different possibilities are 
labeled in Table~\ref{reflection_table_full} as $R_{\eta_T}$, $R_{\eta_C}$, and $R_{\eta_S}$, respectively.
For the symmetry classes that contain two internal symmetries (i.e., classes BDI, CI, CII, and DIII), 
the different (anti-)commutation relations are denoted by $R_{\eta_T \eta_C}$.
Hence, reflection symmetries together with the three internal symmetries
define a total of 27  symmetry classes, see Table~\ref{reflection_table_full}.

The classification of reflection-symmetry-protected semimetals and nodal superconductors depends
on the symmetry class and on the codimension of the Fermi surface 
(superconducting nodes) $p = d_{\mathrm{BZ}} - d_{\mathrm{FS}}$, 
 where $d_{\mathrm{BZ}}$ denotes the
spatial dimension (i.e., the dimension of the Brillouin zone) and $d_{\mathrm{FS}}$ is the dimension of the Fermi surface.
Furthermore, the classification depends on how the Fermi surfaces (superconducting nodes) transform
under the reflection symmetry and the internal symmetries. 
In general, one can distinguish three different situations: 
(i) Each Fermi surface is left invariant by both reflection and internal symmetries,
(ii) Fermi surfaces are invariant under reflection symmetry, but are pairwise
related to each other by the internal symmetries,
and (iii) different Fermi surfaces are pairwise related to each other by both reflection and in internal symmetry operations.
In case (i) and case (ii) the Fermi surfaces are located within the refection plane,
whereas in case (iii) the Fermi surfaces are positioned away from the reflection plane.
For brevity we focus here on only case (i) and case (ii). Case (iii) has been discussed extensively in Refs.~\cite{ChiuSchnyder14} and~\cite{morimotoFurusakiPRB14}.

\subsection{Fermi surfaces at high-symmetry points within mirror plane}

First we consider case (i), i.e., Fermi surfaces that are invariant under both reflection and internal symmetries. These Fermi surfaces
are located within the reflection plane and at high-symmetry points in the Brillouin zone, that is, at time-reversal invariant momenta.
For this case the classification of reflection symmetric Fermi points can be inferred from the classification
of reflection-symmetry-protected insulators by means of a dimensional reduction procedure~\cite{ChiuSchnyder14}. 
Namely, the surface states of  reflection symmetric  $d_{\mathrm{BZ}}$-dimensional topological insulators
can be viewed as reflection-symmetry-protected Fermi points in $d_{\mathrm{BZ}} -1$ dimensions. 
From this it follows that the topological classification of reflection symmetric Fermi points 
at time-reversal invariant momenta  with $d_{\mathrm{FS}} = 0$ is obtained
from the classification of reflection symmetric insulators by the dimensional shift $d_{\mathrm{BZ}} \to d_{\mathrm{BZ}} -1$, see Table~\ref{reflection_table_full}.
This logic also works for Femi surfaces with $d_{\mathrm{FS}} > 0$, if their stability is guaranteed
by an $M\mathbb{Z}$  or $2 M\mathbb{Z}$ topological number.
However, $\mathbb{Z}_2$ and $M\mathbb{Z}_2$ topological numbers ensure only the stability of
Femi points, i.e., Fermi surfaces with $d_{\mathrm{FS}} = 0$.
Derivations based on Clifford analysis and K theory~\cite{Sato_Crystalline_PRB14,ChiuSchnyder14}
corroborate these findings.
The classification of reflection-symmetry-protected Fermi surfaces at high symmetry points of the Brillouin zone
is presented in Table~\ref{reflection_table_full}, where the first row indicates the codimension $p$ of the Fermi surface.
We note that the classification is eight-fold periodic in $p$.

%%%%%%%%%%%%%%%%%%%%%%%%% TABLE
\begin{table*}[thp!]
\caption{
Topological classification of reflection-symmetry-protected semimetals and nodal superconductors.
The first and second rows specify the codimension $p=d_{\mathrm{BZ}}-d_{\mathrm{FS}}$ of the 
reflection symmetric Fermi surfaces (superconducting nodes) at high-symmetry points and away from high-symmetry points of the Brillouin zone, respectively. 
The third row shows the classification of reflection symmetric insulators and fully gapped superconductors~\cite{chiuPRB13,morimotoPRB13}.  
The first column indicates whether the reflection operator $R$ commutes (``$R_+$", ``$R_{++}$") or anti-commutes (``$R_-$", ``$R_{--}$") with the global symmetries.
``$R_{+-}$" and ``$R_{-+}$" denote the case where $R$ commutes with one of the global symmetries but anti-commutes with the other one.
The second column lists the global  symmetry classes (using the ÒCartan nomenclatureÓ~\cite{SchnyderAIP}), which are distinguished by the presence or absence of time-reversal symmetry, 
particle-hole symmetry, and chiral symmetry.
}
\vspace{0.2cm}
\label{reflection_table_full}
\resizebox{\linewidth}{!}{%
\begin{threeparttable}
\begin{tabular}{|c|c|cccccccc|}
\hline
 &  
  $\begin{array}{c}
\mbox{FS in mirror plane} \\
\mbox{ at high-sym.\ point }
\end{array}$
 & $p$=7  & $p$=8 & $p$=1 & $p$=2 & $p$=3 & $p$=4 & $p$=5 & $p$=6    \\
  \cline{2-10}
    $\begin{array}{c}
\mbox{ Reflection } \\
\mbox{ \phantom{a} } 
\end{array}$  
   &  
     $\begin{array}{c}
\mbox{FS in mirror plane} \\
\mbox{ off high-sym.\ point }
\end{array}$  
& $p$=1   & $p$=2 & $p$=3 & $p$=4 & $p$=5 & $p$=6 & $p$=7 & $p$=8     \\
    \cline{2-10}
  & top.\ insul.\ and top.\ SC  & $d_{\mathrm{BZ}}$=8 & $d_{\mathrm{BZ}}$=1 & $d_{\mathrm{BZ}}$=2 & $d_{\mathrm{BZ}}$=3 & 
$d_{\mathrm{BZ}}$=4 & $d_{\mathrm{BZ}}$=5 & $d_{\mathrm{BZ}}$=6 & $d_{\mathrm{BZ}}$=7   \\
\hline
\hline
 $R$ &  A  & 0  & $M\bZ$ & 0 & $M\bZ$ & 0 & $M\bZ$ & 0 & $M\bZ$                \\
 $R_+$ &  AIII & $M\bZ$    & 0 & $M\bZ$ & 0 & $M\bZ$ & 0 & $M\bZ$ & 0           \\ 
$R_-$ &  AIII  & 0  &  $M\bZ\oplus\mathbb{Z}$ & 0 & $M\bZ\oplus\mathbb{Z}$ & 0 & $M\bZ\oplus\mathbb{Z}$ & 0 & $M\bZ\oplus\mathbb{Z}$            \\ 
\hline
\multirow{8}{*}{$R_+$,$R_{++}$} & AI  & $M\bZ_2^{\textrm{a,b}}$    & $M\bZ$ & 0 & 0 & 0 & $2M\bZ$ & 0 & $M\bZ_2^{\textrm{a,b}}$      \\
  & BDI & $M\bZ_2^{\textrm{a,b}}$ & $M\bZ_2^{\textrm{a,b}}$ & $M\bZ$ & 0 & 0 & 0 & $2M\bZ$ & 0         \\
  & D & 0 & $M\bZ_2^{\textrm{a,b}}$ & $M\bZ_2^{\textrm{a,b}}$ & $M\bZ$ & 0 & 0 & 0 & $2M\bZ$          \\
  & DIII & $2M\bZ$    & 0 & $M\bZ_2^{\textrm{a,b}}$ & $M\bZ_2^{\textrm{a,b}}$ & $M\bZ$ & 0 & 0 & 0       \\
  & AII  & 0 & $2M\bZ$ & 0 & $M\bZ_2^{\textrm{a,b}}$ & $M\bZ_2^{\textrm{a,b}}$ & $M\bZ$ & 0 & 0     \\
  & CII & 0  & 0 & $2M\bZ$ & 0 & $M\bZ_2^{\textrm{a,b}}$ & $M\bZ_2^{\textrm{a,b}}$ & $M\bZ$ & 0       \\
  & C  & 0  & 0 & 0 & $2M\bZ$ & 0 & $M\bZ_2^{\textrm{a,b}}$ & $M\bZ_2^{\textrm{a,b}}$ & $M\bZ$        \\
  & CI  & $M\bZ$   & 0 & 0 & 0 & $2M\bZ$ & 0 & $M\bZ_2^{\textrm{a,b}}$ & $M\bZ_2^{\textrm{a,b}}$  
   \\
\hline
\multirow{8}{*}{$R_-$,$R_{--}$} & AI & 0   & 0 & 0 & $2M\bZ$ & 0 & $T\bZ_2^{\textrm{a,b,c}}$ &  $\bZ_2^{\textrm{a,b}}$ & $M\bZ$      \\
  & BDI & $M\bZ$     & 0  & 0 & 0 & $2M\bZ$ & 0 & $T\bZ_2^{\textrm{a,b,c}}$ &  $\bZ_2^{\textrm{a,b}}$      \\
  & D &  $\bZ_2^{\textrm{a,b}}$  &  $M\bZ$   &  0  & 0 & 0 & $2M\bZ$ & 0 & $T\bZ_2^{\textrm{a,b,c}}$        \\
  & DIII  & $T\bZ_2^{\textrm{a,b,c}}$  &  $\bZ_2^{\textrm{a,b}}$ & $M\bZ$   &  0  & 0 & 0 & $2M\bZ$ & 0       \\
  & AII & 0 & $T\bZ_2^{\textrm{a,b,c}}$ &  $\bZ_2^{\textrm{a,b}}$ & $M\bZ$   &  0  & 0 & 0 & $2M\bZ$     \\
  & CII  & $2M\bZ$  & 0 & $T\bZ_2^{\textrm{a,b,c}}$ &  $\bZ_2^{\textrm{a,b}}$ & $M\bZ$   &  0  & 0 & 0        \\
  & C & 0    & $2M\bZ$ & 0 & $T\bZ_2^{\textrm{a,b,c}}$ &  $\bZ_2^{\textrm{a,b}}$ & $M\bZ$   &  0  & 0         \\
  & CI & 0     & 0 & $2M\bZ$ & 0 & $T\bZ_2^{\textrm{a,b,c}}$ &  $\bZ_2^{\textrm{a,b}}$ & $M\bZ$   &  0    
   \\
\hline
$R_{-+}$ & BDI, CII  & 0 & $2\bZ$ & 0 & $2M\bZ$ & 0 & $2\bZ$ & 0 & $2M\bZ$        \\
$R_{+-}$  & DIII, CI & 0  & $2M\bZ$ & 0 & $2\bZ$ & 0 & $2M\bZ$ & 0 & $2\bZ$     \\
\hline
$R_{+-}$  & BDI & $M\bZ_2\oplus \bZ_2^{\textrm{a,b}}$    & $M\bZ\oplus\bZ$ & 0 & 0 & 0 &  $2M\bZ\oplus 2\bZ$ & 0 & $M\bZ_2\oplus \bZ_2^{\textrm{a,b}}$      \\
$R_{-+}$ & DIII  & 0  & $M\bZ_2\oplus \bZ_2^{\textrm{a,b}}$ & $M\bZ_2\oplus \bZ_2^{\textrm{a,b}}$    & $M\bZ\oplus\bZ$ & 0 & 0 & 0 &  $2M\bZ\oplus 2\bZ$     \\
$R_{+-}$  & CII  & 0  &   $2M\bZ\oplus 2\bZ$ & 0 & $M\bZ_2\oplus \bZ_2^{\textrm{a,b}}$ & $M\bZ_2\oplus \bZ_2^{\textrm{a,b}}$    & $M\bZ\oplus\bZ$ & 0 & 0       \\
$R_{-+}$  & CI  & 0   & 0 & 0    & $2M\bZ\oplus 2\bZ$ & 0 & $M\bZ_2\oplus \bZ_2^{\textrm{a,b}}$ & $M\bZ_2\oplus \bZ_2^{\textrm{a,b}}$    & $M\bZ\oplus\bZ$       \\
\hline
\hline
\end{tabular}
 \begin{tablenotes}
  \item[${}^{\textrm{a}}$]    $\bZ_2$ and $M\bZ_2$ invariants only protect Fermi surfaces of dimension zero (i.e., Fermi points with $d_{\mathrm{FS}}=0$) at high-symmetry points of the Brillouin zone. 
        \item[${}^{\textrm{b}}$]   Fermi surfaces located within the mirror plane but away from high symmetry points  cannot be protected by a $\bZ_2$ or $M\bZ_2$  topological number. However, the nodal system can exhibit zero-energy surface states 
        that are protected by a $\bZ_2$ or $M\bZ_2$ topological number.  
                        \item[${}^{\textrm{c}}$]   For topological semimetals (nodal topological superconductors) the presence of translation symmetry is always  assumed.  Therefore, one does not need to distinguish between $T\bZ_2$ and $\bZ_2$ invariants for these gapless topological materials.   
    \end{tablenotes}
    \end{threeparttable}
}
\end{table*}
%%%%%%%%%%%%%%%%%%%%%%%%% TABLE

\subsection{Fermi surfaces within mirror plane but off high-symmetry points}
\label{Sec22}

Second, we review case (ii), i.e., the classification of reflection symmetric Fermi surfaces that
are located away from high-symmetry points of the Brillouin zone.
These Fermi surfaces  transform pairwise into each other by the global (i.e., internal) symmetries, which relate ${\bf k} \to - {\bf k}$.
Using an analysis based on the minimal-Dirac-Hamiltonian method, it
was show in Ref.~\cite{ChiuSchnyder14} that only $M \mathbb{Z}$
and  $2 M \mathbb{Z}$ topological numbers can ensure the stability of reflection symmetric Fermi surfaces off high-symmetry points.
$\mathbb{Z}_2$ and $M \mathbb{Z}_2$ invariants, on the other hand, do not give rise
to stable Fermi surfaces.
However, as illustrated in Sec.~\ref{secDiracSemimetal},
$\mathbb{Z}_2$ or $M \mathbb{Z}_2$  invariants 
may nevertheless lead to protected zero-energy surface states at time-reversal-invariant momenta of the surface Brillouin zone. 
The  classification of reflection symmetric Fermi surfaces located away from high-symmetry points is summarized
in Table~\ref{reflection_table_full}.
As above, we note that 
due to Bott's periodicity theorem~\cite{Bott1970353},
the classification scheme exhibits an eight-fold periodicity in $p$.
We observe that  the classification of reflection-symmetric Fermi surfaces located away from high symmetry points with codimension $p$ is  related to the classification of reflection-symmetric insulators with spatial dimension $d_{\mathrm{BZ}} = p-1$ .

\section{Topological Dirac semimetal with reflection and rotation symmetry}
\label{secDiracSemimetal}

In this section we discuss a tight-binding model describing a topological semimetal which exhibits time-reversal symmetry as well as
reflection and rotation symmetries. This model is inspired by the Dirac material Na$_3$Bi, whose three-dimensional bulk Dirac cones
are protected by reflection symmetry. Using angle-resolved photoemission experiments,
topological surface modes have recently been observed on
the (100) surface of Na$_3$Bi~\cite{Xu18122014}.

\subsection{Model Hamiltonian}

We consider the following cubic-lattice Hamiltonian describing a four-band semimetal with two Dirac points
\begin{subequations} \label{semimetalHam}
\begin{equation} \label{HamRealSpace}
\hat{H}=\frac{1}{2}\sum_{\vec{x}}\Big[ C^\dagger_{\vec{x}+\hat{x}}(\tau_z s_0 + i \tau_x s_x) C^{}_{\vec{x}}+C^\dagger_{\vec{x}+\hat{y}}(\tau_z s_0+ i \tau_x s_y)C^{}_{\vec{x}} +C^\dagger_{\vec{x}+\hat{z}}\tau_z s_0 C^{}_{\vec{x}} + m C^\dagger_{\vec{x}}\tau_z s_0C_{\vec{x}}^{} \Big]+\mathrm{h.c.},
\end{equation}
with the spinor $C_{\vec{x}}=[c_{1\uparrow}(\vec{x}),c_{1\downarrow}(\vec{x}),c_{2\uparrow}(\vec{x}),c_{2\downarrow}(\vec{x})]$, 
where $c_{n \sigma} (\vec{x} )$ denotes the electron annihilation operator with orbital index $n$ and spin $\sigma$.
The two sets of Pauli matrices $\tau_\alpha$ and $s_\alpha$ operate in spin space and orbital space, respectively. 
In momentum space Hamiltonian~\eqref{HamRealSpace} takes diagonal form
$\mathcal{H} =  \sum_{\bf k} C^{\dag}_{\bf k} H ( {\bf k} ) C^{\ }_{\bf k}$,  
with $C^{\ }_{\bf k}$ the Fourier-transformed spinor, and
 \begin{eqnarray} \label{HamKspace}
H({\bf k})=\sin k_x \tix s_x+ \sin k_y \tix s_y + ( \cos k_x + \cos k_y + \cos k_z-m) \tiz s_0  .
\end{eqnarray}
\end{subequations}
For concreteness we set $m=2$. With this choice, the bulk Dirac cones of $H({\bf k})$ are positioned at
$\bk_\pm=(0,0,\pm \pi/2)$, i.e., they are located on the $k_z$ axis just as in the Dirac material Na$_3$Bi.
The cubic-lattice Hamiltonian~\eqref{HamKspace} possesses a rotation symmetry of order four about the
$z$ axis. This $C_4$ rotation symmetry acts on $H ( {\bf k} )$ as 
\begin{eqnarray} \label{C4RotSym}
R^{-1}_{C_4}H(- k_y, k_x,k_z)R^{\ }_{C_4}=H(k_x, k_y, k_z),
\end{eqnarray}
where $R_{C_4}=\tau_z(s_0+is_z)/\sqrt{2}$ denotes the  $C_4$ rotation operator.
The Dirac semimetal~\eqref{semimetalHam} also preserves time reversal symmetry, Eq.~\eqref{eqTRS},  
with the time-reversal operator $T=\tau_0 s_y \ck$. 
Since $T^2 = - \mathbbm{1}$,
Hamiltonian~\eqref{semimetalHam} belongs to symmetry class~AII. The codimension of the Dirac cones of $H({\bf k})$, which are located
away from the high symmetry points in the Brillouin zone, is $p=3$. 
Hence, according to the classification of Ref.~\cite{ChiuSchnyder14} (see Table~1 in Ref.~\cite{ChiuSchnyder14}), the Dirac points are not protected by time-reversal symmetry, even though
a binary $\mathbb{Z}_2$ invariant can be defined for this symmetry class. (As we will show below,  this $\mathbb{Z}_2$ invariant leads to gapless surface states at time-reversal-invariant momenta of the surface Brillouin zone.) 
Indeed, we find that the bulk Dirac cones can be gapped out by the 
mass term $\sin k_z \tiy s_0$, which preserves time reversal symmetry. 
However, the mass term  $\sin k_z \tiy s_0$ breaks the $C_4$ rotation symmetry~\eqref{C4RotSym}.
In general, we find that the only possible gap opening terms are $f_1(k_z)\tiy s_0$ and $f_2(k_z)\tix s_z$, 
since these are the terms that anti-commute with $H ( {\bf k} )$.
Here,  $f_1(k_z)$ and $f_2(k_z)$ represent $k_z$ dependent masses. 
We find that these two gap opening terms break the $C_4$ rotation symmetry~\eqref{C4RotSym}, since they anti-commute
with $R_{C_4}$. As a consequence, the two Dirac cones of Hamiltonian~\eqref{semimetalHam} are protected against
gap opening by the $C_{4}$ rotation symmetry~\eqref{C4RotSym}.

\subsection{Surface states}

\begin{figure}
\begin{center}
\includegraphics[clip,width=1.0\columnwidth]{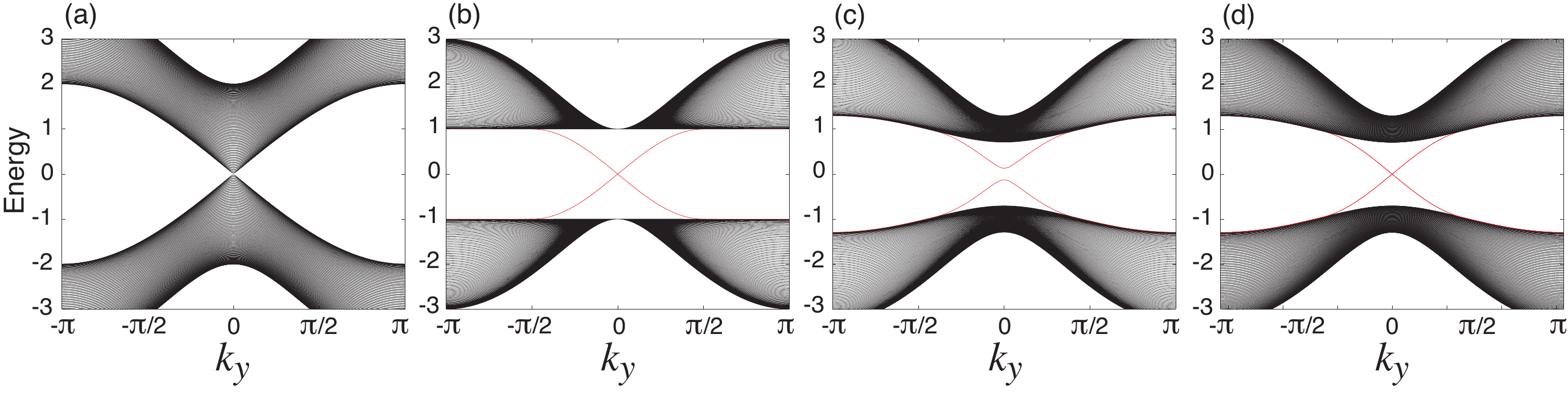}
\end{center}
  \caption{\label{surfacBandStruct2D} 
(a) Surface band structure of the Dirac semimetal~\eqref{semimetalHam} for a (100) slab as a function of  surface momentum $k_y$ with fixed $k_z = \pi/2$. (b) Same as panel (a) but for $k_z = 0$. The gapless surface state at $k_y=0$ is protected by a $\mathbb{Z}_2$ invariant.
(c) Band structure at the (100) surface of the Dirac semimetal~\eqref{modifiedHam} as a function of $k_y$ with $k_z= \pi /4$. In the absence of reflection symmetry~\eqref{reflectSym} and chiral symmetry~\eqref{chiralSym} the surface state (red trace) is gapped.
(d) Surface band structure of the Dirac semimetal~\eqref{semimetalHam} (i.e., with reflection symmetry and chiral symmetry) as a function of $k_y$ with $k_z = \pi /4$. 
}   
\end{figure}

As discussed above and in Ref.~\cite{ChiuSchnyder14},  class AII Dirac points with codimension $p=3$ that are located away from high-symmetry points of the Brillouin zone are unstable in the 
absence of crystal lattice symmetries, even though a binary $\mathbb{Z}_2$ index can be defined for this symmetry class. This 
$\mathbb{Z}_2$ invariant does not guarantee the stability of  Dirac points, however it gives rise to protected gapless surface states
at time-reversal invariant momenta of the surface Brillouin zone. For the Dirac semimetal~\eqref{semimetalHam} these gapless modes appear at $(k_y, k_z)=(0,0)$ of the (100) surface Brillouin zone, see Fig.~\ref{surfacBandStruct2D}(b). For other surface momenta, the surface states are in general gapped.
Similarly, rotation symmetry~\eqref{C4RotSym} can protect surface states only at time-reversal-invariant momenta of the surface Brillouin zone, but not away from these points.

%%%%%%%%%%%%%%%%%%%%%%%%%%%%
%%%%%%%%%%%%%%%%%%%%%%%%%%%%

\begin{figure}
\centering
\begin{center}
\includegraphics[clip,width=0.7\columnwidth]{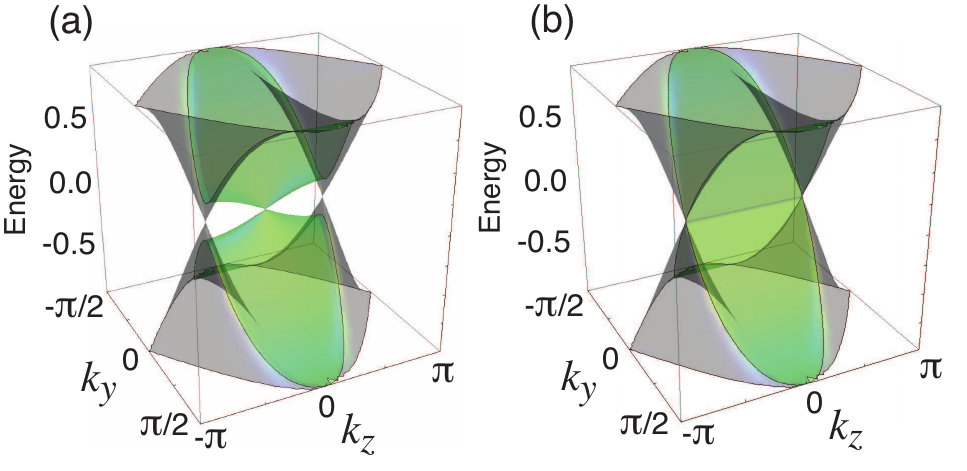}
\end{center}
  \caption{\label{surfacBandStruct3D} 
(a)
Surface spectrum of the Dirac semimetal~\eqref{modifiedHam} for the (100) face as a function of  surface momenta $k_y$ and $k_z$. 
The surface states and bulk states are colored in green and gray, respectively. In the absence of reflection symmetry~\eqref{reflectSym}
and chiral symmetry~\eqref{chiralSym} the surface state is gapped except at $(k_y, k_z)=(0,0)$.
(b) Surface spectrum of the Dirac semimetal~\eqref{semimetalHam}, i.e., in the presence of reflection symmetry and chiral symmetry.
The gaplessness of the Fermi arc is guaranteed by the mirror winding number~\eqref{eqMirrorWinding}.
}   
\end{figure}

To illustrate this, we consider Hamiltonian~\eqref{semimetalHam} with periodic boundary conditions along the $z$-axis and open boundary conditions along the $x$- and $y$-directions. We observe that model Hamiltonian~\eqref{semimetalHam} exhibits besides rotation symmetry \eqref{C4RotSym} an \emph{accidental} reflection symmetry 
\begin{eqnarray} \label{reflectSym}
R_y^{-1}H(k_x,-k_y,k_z) R_y&= H(k_x,k_y, k_z), 
\end{eqnarray}
with $R_y=\tiz s_y$,
and an \emph{accidental} chiral symmetry  
\begin{eqnarray} \label{chiralSym}
H( \bk)S&= - S H(\bk),
\end{eqnarray}
with $S=\tix s_z$.
In order to break these accidental symmetries we need to add to the Hamiltonian the term $+g\sin k_z \tau_x s_z$ on the (100) and  ($\bar{1}$00) surfaces,
 and the term $-g\sin k_z \tau_x s_z$ on the (010) and  (0$\bar{1}$0) surfaces. That is, we consider
\begin{eqnarray} \label{modifiedHam}
\widetilde{H} (k_z ) 
=
\hat{H} ( k_z) + g \sum_{\vec{x}_{\perp}} C^{\dag}_{\vec{x}_{\perp}} 
\sin k_z \tau_x s_z
\left[
\delta_{\vec{x}_{\perp},\vec{x}^{(100)}_{\perp}} 
+ \delta_{\vec{x}_{\perp},\vec{x}^{(\bar{1}00)}_{\perp}} 
- \delta_{\vec{x}_{\perp},\vec{x}^{(010)}_{\perp}}  - \delta_{\vec{x}_{\perp},\vec{x}^{(0 \bar{1} 0)}_{\perp}} 
\right]
C^{\ }_{\vec{x}_{\perp}} ,
\end{eqnarray}
where $\hat{H} ( k_z)$ is obtained from Eq.~\eqref{HamRealSpace} by Fourier transforming the $z$-coordinate 
and the sum in Eq.~\eqref{modifiedHam} is over a collection of one-dimensional chains which are oriented along the $z$-axis and labeled by $\vec{x}_\perp$. Hamiltonian \eqref{modifiedHam} is symmetric under time-reversal symmetry and
rotation symmetry~\eqref{C4RotSym}, but breaks the accidental  symmetries \eqref{reflectSym} and~\eqref{chiralSym}.
Fig.~\ref{surfacBandStruct2D}(c)  shows the (100) surface spectrum of Hamiltonian~\eqref{modifiedHam} with $g= 0.2$ as
a function of surface momentum $k_y$ for $k_z= \pi / 4$. 
Fig.~\ref{surfacBandStruct3D}(a) displays the (100) surface spectrum as a function of both surface momenta $k_y$ and $k_z$.
We observe that the surface state is gapped, except at $(k_y, k_z)=(0,0)$, where it is protected by time-reversal symmetry (i.e., the aforementioned $\mathbb{Z}_2$ invariant).

On the other hand, if we consider a semimetal which possesses also symmetries~\eqref{reflectSym} and~\eqref{chiralSym}
[i.e., Hamiltonian~\eqref{semimetalHam} instead of Hamiltonian~\eqref{modifiedHam}], then the surface states are gapless along a one-dimensional arc that connects the two projected Dirac nodes $\bk_\pm=(0,0,\pm \pi/2)$, see Fig.~\ref{surfacBandStruct2D}(d) and
Fig.~\ref{surfacBandStruct3D}(b).
Since the reflection operator $R_y$ anti-commutes with the
time-reversal operator $T=\tau_0 s_y \ck$ and  with the effective particle-hole operator
$C = T^{-1} S = i \tau_x s_x \mathcal{K}$, Hamiltonian~\eqref{semimetalHam}  can be viewed
as a member of symmetry class DIII with $R_{--}$ in Table~\ref{reflection_table_full}.
Hence, according to the classification scheme of Sec.~\ref{Sec22} the Dirac nodes of Eq.~\eqref{semimetalHam} are
protected by an $M\mathbb{Z}$ invariant, i.e., a mirror winding number. Moreover, Hamiltonian~\eqref{semimetalHam} exhibits Fermi arc surface states, which
are protected by this mirror winding number. This mirror invariant is defined in terms of a one-dimensional integral
along a contour that lies within the mirror plane $k_y=0$. Within the $k_y=0$ mirror plane
the Hamiltonian can be block diagonalized with respect to $R_y$. For the block with mirror eigenvalue 
$R_y=+1$ the mirror winding number takes the form
\begin{eqnarray} \label{eqMirrorWinding}
n^{+}_{\mathcal{C}}Ê =\frac{i}{2\pi}\int_{\mathcal{C}}\Tr (q^*dq), 
\quad
\textrm{with}
\quad
q=\frac{\sin k_x - i M(k_x)}{\sin^2 k_x + M^2(k_x)}
\end{eqnarray}
and $M(k_x)=\cos k_x + \cos k_z -1$, where
$\mathcal{C}$ denotes a contour that lies within the mirror plane. Choosing the contour $\mathcal{C}$
to be parallel to the $k_x$-axis with $k_y=0$ and $k_z$ a fixed parameter we find that  
\bee 
n^{+}Ê(k_z)=\left\{ 
  \begin{array}{r l}
    -1,& \, \left| k_z \right| < \pi/2 \\
    0,&   \,  \left| k_z \right| > \pi/2
  \end{array} \right.  .
\ee	 
This indicates that there exists a gapless Fermi arc state at $k_y=0$ on the (100) surface,  
which is in agreement with the spectrum shown in Fig.~\ref{surfacBandStruct3D}(b).

%%%%%%%%%%%%%%%%%%%%%%%%%%%%
%%%%%%%%%%%%%%%%%%%%%%%%%%%%

\subsection{The role of translation symmetry}

While each Dirac point of a topological semimetal is stable against deformations that are local in momentum space,
the Dirac points are not protected against commensurate perturbations, such as charge-density wave modulations, which connect Dirac points with opposite momenta. However, these deformations are forbidden, if we impose besides internal and crystal point-group symmetries, also translation symmetry.

To illustrate the instability of the Dirac points against charge-density wave perturbations, let us consider a low-energy description
of semimetal~\eqref{semimetalHam}. Expanding Eq.~\eqref{semimetalHam} around the two Dirac points $\bk_\pm=(0,0,\pm \pi/2)$, we obtain the
following low-energy effective Hamiltonian
\bee
\hat{H}_{\textrm{eff}}=
\sum_{k_x, k_y, k_\Delta}
\bma 
C^\dagger_{{\bf K}_{+}} & C^\dagger_{{\bf K}_{-}}
\ema
\bma 
h_+ & 0 \\
0 & h_- \\
\ema
\bma
C_{{\bf K}_{+}} \\
C_{{\bf K}_{-}} \\
\ema,
\ee 
where  ${\bf K}_\pm =(k_x,k_y, k_\Delta \pm \pi/2 )$ and $h_\pm =  k_x \tix s_x+ k_y \tix s_y \mp k_\Delta  \tiz s_0 $. 
Hamiltonian $\hat{H}_{\textrm{eff}}$ is symmetric under the $C_4$ rotation symmetry~\eqref{C4RotSym}, which transforms $C_{{\bf K}_\pm}\rightarrow \tau_z(s_0+is_z)C_{\widetilde{\bf K}_{\pm}}/\sqrt{2} $, and the time reversal symmetry~\eqref{eqTRS}, which transforms $C_{{\bf K}_\pm}\rightarrow \tau_0 s_y\mathcal{K} C_{-{\bf K}_{\pm}}$. 
We find that the two Dirac cones can be gapped out by the 
charge-density wave modulation
\bee
\hat{D}=
\sum_{k_x, k_y, k_\Delta}
\bma 
C^\dagger_{{\bf K}_{+}} & C^\dagger_{{\bf K}_{-}}
\ema
\bma 
0 & \tiz s_0 \\
\tiz s_0 & 0 \\ 
\ema
\bma
C_{{\bf K}_{+}} \\
C_{{\bf K}_{-}} \\
\ema,
\ee
which preserves rotation symmetry~\eqref{C4RotSym}, but
breaks translation symmetry.
We conclude that in the absence of translation symmetry, the Dirac cones of semimetals are unstable
against charge-density wave-type perturbations.

\subsection{Implications for Na$_3$Bi}
\label{secImplicationNa3Bi}

Albeit Na$_3$Bi has hexagonal $P6_3/mmc$ crystal structure rather than cubic one, its topological and symmetry properties are  similar to the model Hamiltonian~\eqref{modifiedHam}. Instead of $C_4$ rotation symmetry, Na$_3$Bi exhibits a $C_3$ rotation symmetry, which protects
the two gapless bulk Dirac cones. Using a downfolding procedure, it is possible to derive from   DFT band structure calculations a low-energy  hexagonal-lattice tight-binding model for Na$_3$Bi. This tight-binding model is expected to exhibit  similar symmetry properties
and the same topological features as our tight-binding Hamiltonian~\eqref{modifiedHam}. Hence, this suggests that Na$_3$Bi supports
Fermi arc surface states, which are gapped except at time-reversal invariant points of the surface Brillouin zone.
This expectation is confirmed by recent angle-resolved photoemission experiments~\cite{xuLiuHasanArxiv13,Xu18122014}.

\section{Summary and discussion}

In this paper, we have reviewed the topological classification of reflection-symmetry-protected semimetals and nodal superconductors. 
This classification scheme depends on (i) the internal symmetry properties of the Hamiltonian, (ii) whether the reflection symmetry commutes or anti-commutes with the internal symmetries,
(iii) the codimension of the Fermi surface (superconducting node) $p= d_{\mathrm{BZ}} - d_{\mathrm{FS}}$, 
and (iv) the transformation properties of the Fermi surface (superconducting node) under the action of the reflection and internal symmetries.
Reflection symmetry together with the internal (i.e., non-spatial) symmetries define a total of 27 symmetry classes. 
The result of the classification scheme is presented in Table~\ref{reflection_table_full}.
The stability of the Fermi surface is ensured by the conservation of a topological invariant 
that is defined as an integral along a contour surrounding the Fermi surface.
These topological numbers can be Chern or winding numbers (indicated by ``$\mathbb{Z}$" in Table~\ref{reflection_table_full}), binary invariants  (indicated by ``$\mathbb{Z}_2$"), mirror Chern or mirror winding numbers (indicated by ``$M\mathbb{Z}$"), or mirror binary invariants (indicated by ``$M\mathbb{Z}_2$").

To illustrate the usefulness of the classification Table~\ref{reflection_table_full},  we have discussed in Sec.~\ref{secDiracSemimetal}
an example of a  topological semimetal with time-reversal symmetry and rotation symmetry. This model Hamiltonian represents
a cubic lattice version of the hexagonal Dirac material Na$_3$Bi. We determined the surface state spectrum
of this Dirac model system and discussed its topological properties. We found that the (100) surface supports Fermi arc surface states, which are gapped except at the $\bar{\Gamma}$ point of the surface Brillouin zone, see Fig.~\ref{surfacBandStruct3D}(a). This is in agreement with recent angular-resolved photoemission experiments on the (100) surface of Na$_3$Bi~\cite{xuLiuHasanArxiv13,Xu18122014}.

\ack
The authors thank Wei-Feng Tsai for useful discussions. 
The support of the Max-Planck-UBC Centre for Quantum Materials is gratefully acknowledged.

\section*{References}

\bibliography{references}

\providecommand{\newblock}{}
\begin{thebibliography}{10}
\expandafter\ifx\csname url\endcsname\relax
  \def\url#1{{\tt #1}}\fi
\expandafter\ifx\csname urlprefix\endcsname\relax\def\urlprefix{URL }\fi
\providecommand{\eprint}[2][]{\url{#2}}

\bibitem{Schnyder2008gf}
Schnyder A~P, Ryu S, Furusaki A and Ludwig A~W~W 2008 {\em Phys. Rev. B\/} {\bf
  78} 195125

\bibitem{Kitaev}
Kitaev A 2009 {\em AIP Conf. Proc.\/} {\bf 1134} 22

\bibitem{SchnyderAIP}
Schnyder A~P, Ryu S, Furusaki A and Ludwig A~W~W 2009 {\em AIP Conf. Proc.\/}
  {\bf 1134} 10

\bibitem{Ryu2010ten}
Ryu S, Schnyder A~P, Furusaki A and Ludwig A~W~W 2010 {\em New J. Phys.\/} {\bf
  12} 065010

\bibitem{Chiu_nontrivial_surface}
{Chiu} C~K 2014 {\em ArXiv e-prints\/} (\textit{Preprint} \eprint{1410.1117})

\bibitem{teoPRB08}
Teo J~C~Y, Fu L and Kane C~L 2008 {\em Phys. Rev. B\/} {\bf 78} 045426

\bibitem{Fu2011uq}
Fu L 2011 {\em Phys. Rev. Lett.\/} {\bf 106} 106802

\bibitem{slaberNatPhys13}
Slager R~J, Mesaros A, Juricic V and Zaanen J 2013 {\em Nat. Phys.\/} {\bf 9}
  98

\bibitem{uenoPRL13}
Ueno Y, Yamakage A, Tanaka Y and Sato M 2013 {\em Phys. Rev. Lett.\/} {\bf 111}
  087002

\bibitem{zhangPRL13}
Zhang F, Kane C~L and Mele E~J 2013 {\em Phys. Rev. Lett.\/} {\bf 111} 056403

\bibitem{benalcazar2013}
Benalcazar W~A, Teo J~C~Y and Hughes T~L 2014 {\em Phys. Rev. B\/} {\bf 89}(22)
  224503

\bibitem{Teo_hughes_2013}
Teo J~C~Y and Hughes T~L 2013 {\em Phys. Rev. Lett.\/} {\bf 111}(4) 047006

\bibitem{Turner:2012bh}
Turner A~M, Zhang Y, Mong R~S~K and Vishwanath A 2012 {\em Phys. Rev. B\/} {\bf
  85} 165120

\bibitem{HughesPRB11}
Hughes T~L, Prodan E and Bernevig B~A 2011 {\em Phys. Rev. B\/} {\bf 83} 245132

\bibitem{chiuPRB13}
Chiu C~K, Yao H and Ryu S 2013 {\em Phys. Rev. B\/} {\bf 88}(7) 075142

\bibitem{morimotoPRB13}
Morimoto T and Furusaki A 2013 {\em Phys. Rev. B\/} {\bf 88}(12) 125129

\bibitem{Sato_Crystalline_PRB14}
Shiozaki K and Sato M 2014 {\em Phys. Rev. B\/} {\bf 90}(16) 165114

\bibitem{Tanaka:2012fk}
Tanaka Y, Ren Z, Sato T, Nakayama K, Souma S, Takahashi T, Segawa K and Ando Y
  2012 {\em Nat Phys\/} {\bf 8} 800

\bibitem{Hsieh:2012fk}
Hsieh T~H, Lin H, Liu J, Duan W, Bansil A and Fu L 2012 {\em Nat. Commun.\/}
  {\bf 3} 982

\bibitem{Xu2012}
Xu S~Y, Liu C, Alidoust N, Neupane M, Qian D, Belopolski I, Denlinger J~D, Wang
  Y~J, Lin H, Wray L~A, Landolt G, Slomski B, Dil J~H, Marcinkova A, Morosan E,
  Gibson Q, Sankar R, Chou F~C, Cava R~J, Bansil A and Hasan M~Z 2012 {\em Nat.
  Commun.\/} {\bf 3} 1192

\bibitem{Dziawa2012uq_short}
\mbox{Dziawa} {\it et al\/} P 2012 {\em Nat. Mater.\/} {\bf 11} 1023

\bibitem{OgataJPSJ2011}
Kariyado T and Ogata M 2011 {\em Journal of the Physical Society of Japan\/}
  {\bf 80} 083704

\bibitem{Kariyado_Ca3PbO}
Kariyado T and Ogata M 2012 {\em Journal of the Physical Society of Japan\/}
  {\bf 81} 064701

\bibitem{hsieFuAntiPero14}
Hsieh T~H, Liu J~W and Fu L 2014 {\em Phys. Rev. B\/} {\bf 90} 08112

\bibitem{WanVishwanathSavrasovPRB11}
Wan X, Turner A~M, Vishwanath A and Savrasov S~Y 2011 {\em Phys. Rev. B\/} {\bf
  83}(20) 205101

\bibitem{SchnyderRyuFlat}
Schnyder A~P and Ryu S 2011 {\em Phys. Rev. B\/} {\bf 84}(6) 060504

\bibitem{BrydonSchnyderTimmFlat}
Brydon P~M~R, Schnyder A~P and Timm C 2011 {\em Phys. Rev. B\/} {\bf 84}(2)
  020501

\bibitem{Brydon10}
Schnyder A~P, Brydon P~M~R and Timm C 2011 {\em Phys. Rev. B\/} {\bf 85} 24522

\bibitem{matsuuraNJP13}
Matsuura S, Chang P~Y, Schnyder A~P and Ryu S 2013 {\em New J. Phys.\/} {\bf
  15} 065001

\bibitem{ZhaoWangPRL13}
Zhao Y~X and Wang Z~D 2013 {\em Phys. Rev. Lett.\/} {\bf 110}(24) 240404

\bibitem{ZhaoWangPRB14}
Zhao Y~X and Wang Z~D 2014 {\em Phys. Rev. B\/} {\bf 89}(7) 075111

\bibitem{turnerVishwanathReview}
{Turner} A~M and {Vishwanath} A 2013 {\em ArXiv e-prints\/} (\textit{Preprint}
  \eprint{1301.0330})

\bibitem{ChiuSchnyder14}
Chiu C~K and Schnyder A~P 2014 {\em Phys. Rev. B\/} {\bf 90}(20) 205136

\bibitem{YangNagaosaNatComm14}
Yang B~J and Nagaosa N 2014 {\em Nat. Commun.\/} {\bf 5} 4898

\bibitem{morimotoFurusakiPRB14}
Morimoto T and Furusaki A 2014 {\em Phys. Rev. B\/} {\bf 89}(23) 235127

\bibitem{poYao14}
Chang P~Y, Matsuura S, Schnyder A~P and Ryu S 2014 {\em Phys. Rev. B\/} {\bf
  90}(17) 174504

\bibitem{Yazdani_CdAs}
Jeon S, Zhou B~B, Gyenis A, Feldman B~E, Kimchi I, Potter A~C, Gibson Q~D, Cava
  R~J, Vishwanath A and Yazdani A 2014 {\em Nat. Mater.\/} {\bf 13} 851

\bibitem{Dirac_semimetal_Xi_Dai}
Wang Z, Weng H, Wu Q, Dai X and Fang Z 2013 {\em Phys. Rev. B\/} {\bf 88}(12)
  125427

\bibitem{neupaneDiracHasan}
Neupane M, Xu S~Y, Sankar R, Alidoust N, Bian G, Liu C, Belopolski I, Chang
  T~R, Jeng H~T, Lin H, Bansil A, Chou F and Hasan M~Z 2014 {\em Nat.
  Commun.\/} {\bf 5} 3786

\bibitem{borisenkoPRLCd3As2}
Borisenko S, Gibson Q, Evtushinsky D, Zabolotnyy V, B\"uchner B and Cava R~J
  2014 {\em Phys. Rev. Lett.\/} {\bf 113}(2) 027603

\bibitem{Cd3As2Chen2014}
Liu Z~K, Jiang J, Zhou B, Wang Z~J, Zhang Y, Weng H~M, Prabhakaran D, Mo S~K,
  Peng H, Dudin P, Kim T, Hoesch M, Fang Z, Dai X, Shen Z~X, Feng D~L, Hussain
  Z and Chen Y~L 2014 {\em Nat. Mater.\/} {\bf 13} 677

\bibitem{heLiCd3AsTransport}
He L\ P, Hong X\ C, Dong J\ K, Pan J, Zhang Z, Zhang J and Li S\ Y 2014 {\em
  Phys. Rev. Lett.\/} {\bf 113}(24) 246402

\bibitem{liangOngTransportCd3As2}
Liang T, Gibson Q, Ali M~N, Liu M, Cava R~J and Ong N~P 2014 {\em Nature
  Materials\/} (\textit{Preprint}
  \eprint{http://nmat/journal/vaop/ncurrent/full/nmat4143.html})

\bibitem{Liu21022014}
Liu Z~K, Zhou B, Zhang Y, Wang Z~J, Weng H~M, Prabhakaran D, Mo S~K, Shen Z~X,
  Fang Z, Dai X, Hussain Z and Chen Y~L 2014 {\em Science\/} {\bf 343} 864

\bibitem{Dai_predition_Na3Bi}
Wang Z, Sun Y, Chen X~Q, Franchini C, Xu G, Weng H, Dai X and Fang Z 2012 {\em
  Phys. Rev. B\/} {\bf 85}(19) 195320

\bibitem{xuLiuHasanArxiv13}
{Xu} S~Y, {Liu} C, {Kushwaha} S~K, {Chang} T~R, {Krizan} J~W, {Sankar} R,
  {Polley} C~M, {Adell} J, {Balasubramanian} T, {Miyamoto} K, {Alidoust} N,
  {Bian} G, {Neupane} M, {Belopolski} I, {Jeng} H~T, {Huang} C~Y, {Tsai} W~F,
  {Lin} H, {Chou} F~C, {Okuda} T, {Bansil} A, {Cava} R~J and {Hasan} M~Z 2013
  {\em ArXiv e-prints\/} (\textit{Preprint} \eprint{1312.7624})

\bibitem{Xu18122014}
Xu S~Y, Liu C, Kushwaha S~K, Sankar R, Krizan J~W, Belopolski I, Neupan M, e,
  Bian G, Alidoust N, Chang T~R, Jeng H~T, Huang C~Y, Tsai W~F, Lin H, Shibayev
  P~P, Chou F, Cava R~J and Hasan M~Z 2014 {\em Science\/} (\textit{Preprint}
  \eprint{http://www.sciencemag.org/content/early/2014/12/17/science.1256742.full.pdf})

\bibitem{Bott1970353}
Bott R 1970 {\em Advances in Mathematics\/} {\bf 4} 353

\end{thebibliography}

\end{document}